\documentstyle[PASJadd,psfig]{PASJ95}
\draft 
\begin{document}
\setcounter{page}{1}

\title{X-Ray Observation of the L1157 Dark Cloud Region with ASCA} 
\author{Tae {\sc Furusho},
Noriko {\sc Y. Yamasaki}, and Takaya {\sc Ohashi}, \\
{\it Department of Physics, Tokyo Metropolitan University,
1-1 Minami-Ohsawa,} \\
{\it  Hachioji, Tokyo 192-0397} \\
{\it E-mail(TF): furusho@phys.metro-u.ac.jp}\\[6pt]
Yoshitaka {\sc Saito} \\
{\it Institute of Space and Astronautical Science (ISAS),}\\
{\it 3-1-1 Yoshinodai, Sagamihara, Kanagawa 229-8510}\\
and \\
Wolfgang {\sc Voges} \\
{\it Max-Planck-Institut f\"{u}r Extraterrestrische Physik,
D-85748, Garching, Germany}}

\abst{ ASCA observation of a region containing a Class 0 protostar,
  IRAS 20386+6751 in the L1157 dark cloud, has been carried out.  The
  protostar was not detected, and the 95\% upper limit to the
  luminosity depends on assumed $N_{\rm H}$: $L_{\rm X}(0.5-10\ {\rm
  keV}) < 1.1 \times10^{31}$ erg s$^{-1}$ for $N_{\rm H} =
  1\times10^{23}$ cm$^{-2}$.  A Class I protostar in L1152, IRAS 20353+6742, in the same
  field was also undetected with the upper limit about three times as much
  as the L1157 level. Besides these non detections, nine new X-ray
  sources were detected and spectral analysis was performed for 4
  sources. One object (AXJ 2038+6801) shows a hard spectrum with a
  temperature $kT\sim8$ keV or a power-law photon index $\sim 2.0$ and
  absorbed with $N_{\rm H}\sim 2\times10^{22}$ cm$^{-2}$. Another
  fainter one (AXJ 2036+6800) has a soft spectrum with most of the
  emission falling below 2 keV\@. We examine possible nature of these
  new X-ray sources based on their spectral properties.}

\kword{Stars: pre-main-sequence --- Stars: X-rays --- Stars: individual (L1157)
--- X-rays: spectra --- X-rays: sources}

\maketitle \thispagestyle{headings}

\section{Introduction}

Young stellar objects are mainly classified into 4 classes based on
the slope of their spectral energy distribution (SED) at infrared to
millimeter wavelength (Andr\'e et al.\ 1993); Class 0 and
I protostars, Class II classical T Tauri stars, and Class III
weak-lined T Tauri Stars.  Class 0 protostars are thought to be in an
evolutionary stage prior to Class I and are still in an active
accretion phase.  The typical life time of Class 0 objects is
estimated to be about $ 10^4$ yr and the temperature derived by
fitting the spectrum with a single black body model is less than 100
K\@.  It is also known that well collimated CO outflows are associated
with most of Class 0 protostars (Bachiller 1996).  X-ray emission from
Class I protostars has been discovered by ROSAT and ASCA satellites
(Casanova et al.\ 1995, Koyama et al.\ 1996, Kamata et al.\ 1997,
Ozawa et al.\ 1999); however, not all Class I protostars are X-ray
emitters.  Their spectra are harder than those of T Tauri Stars, and
the X-ray emission sometimes shows time variability like solar
flares. Class I objects detected with ASCA, for instance WL6 and EL29
in the $\rho$ Ophiuchi dark cloud, indicate 2 -- 10 keV luminosities in the
low state to be 2.5 and $3.3\times10^{30}\ {\rm erg\ s}^{-1}$,
respectively (Kamata et al.\ 1997). Their spectra can be fitted with
an absorbed thermal bremsstrahlung model with a temperature
$kT\sim$2--3 keV, and a hydrogen column density $N_{\rm
H}\sim4\times10^{22} {\rm cm}^{-2}$.  The mechanism of the X-ray
emission from protostars, however, has not been clearly understood
yet. Naively, it is likely that  the release of kinetic energy or angular 
momentum of the accreting matter drives both the CO outflow 
and X-ray emission.
Our motivation for this observation is to study in what
evolutionary stage a protostar starts to emit X-rays. The X-ray
emission from a Class 0 protostar, if detected, would provide useful
clues to solve this important problem.

In the L1157 cloud, a well collimated CO outflow is found by Umemoto
et al.\ (1992), which is driven by IRAS 20386+6751 with its low
luminosity ($\approx 11 L_{\solar}$) at a distance of 440 pc.  The
outflow extends over about $5'$ ($\sim0.6$ pc) and the dynamical time
scale is estimated to be 2--3 $\times 10^4$ yr.  The mass of the
central object is $\sim 0.2 M_{\solar}$, lower than the estimated mass
of the surrounding envelope $M_{\rm env}\sim 3 M_{\solar}$ (Gueth et al.\
1997). The fact that most of the mass is in the envelope suggests that
this young object -- hereafter called ``L1157-mm'' -- is still in main
accretion phase.  The SED of the IRAS source, $F_{\nu}(100~\mu{\rm
m})/F_{\nu}(60~\mu{\rm m})=4.0$, also shows the characteristics of the
Class 0 objects (Tafalla, Bachiller 1995).  Although ASCA has
observed several tens of Class 0 protostars, no clear X-ray emission
has been detected from them. Most of the objects lie in crowded
regions, making a clear identification difficult even if there was an
indication of X-ray emission.  Since L1157-mm is relatively isolated,
we have little ambiguity in determining the origin of X-ray emission
to be whether from L1157-mm or from others with the ASCA data.

The GIS field of view includes a Class I protostar in the L1152 dark
cloud, which has been identified as IRAS 20353+6742.  The coordinate of
IRAS 20353+6742 is (R.A., Decl.)$_{\rm J2000}$ = ($20^{\rm h}35^{\rm m}46^{\rm s}$, $67^{\circ}53'2''$), and it is about $ 21'$ away from
L1157-mm.  This source also shows the CO bipolar outflow detected by
Bontemps et al.\ (1996), and the envelope mass is estimated to be
$0.05 M_{\solar}$. The distance is the same as that of L1157-mm and the
bolometric luminosity is $3.3 L_{\solar}$ (Myers et al.\ 1987).

\section{Observation}

We observed L1157-mm with the ASCA satellite (Tanaka et al.\ 1994) on
1998 September 5 for a net exposure time of about 60 ks. The
coordinate of L1157-mm position is (R.A., Decl.)$_{\rm J2000}$ = ($20^{\rm h}39^{\rm m}6^{\rm s}$, $68^{\circ}2'13''$).  ASCA has four identical
X-ray telescopes (XRT) which collect incident X-rays onto two
Solid-state Imaging Spectrometers (SIS0, SIS1) and two Gas Imaging
Spectrometers (GIS2, GIS3)\@.  The field of view of each SIS is
$11'\times11'$ and the covered energy range is 0.4 -- 10 keV with an
energy resolution of about 160 eV (FWHM) at 6 keV (Burke et al.\
1991).  The GIS instrument has a field of view of $40'$ diameter and
covers the energy range from 0.7 keV to 10 keV with an energy
resolution of 0.5 keV (FWHM) at 5.9 keV (Ohashi et al.\ 1996).  During
this observation, the SIS was operated in the 1-CCD mode, and the GIS
was in the normal PH mode with the nominal bit assignments.

\section{Analysis and Results}

\subsection{Upper Limits for L1157-mm and IRAS 20353+6742}


Figure 1 shows GIS images in the soft (0.7 -- 2.0 keV), and hard (2.0 --
10 keV) energy bands. The images are smoothed by a Gaussian function
with a $\sigma$ of $0.5'$. The position of L1157-mm is shown with an
open circle located about $6.5'$ (0.8 pc) east of the field center.
There is no particular X-ray excess corresponding to L1157-mm in
either GIS or SIS image.  We calculated the upper limit of the X-ray
emission from L1157-mm for the GIS data because systematic error of
the SIS background was higher than that of the GIS\@. There are two
different ways of background estimation. One is to take the
symmetrical position of the source with respect to the optical axis,
and the other is using the public GIS blank sky event data which are
already corrected for point sources.  The former is a reasonable
method for background subtraction because the data are taken
simultaneously, but the latter gives higher statistics.  As a result,
both methods gave consistent results within the error at the 90\%
confidence level.  The upper limit for the L1157-mm intensity at the
95\% confidence is estimated to be $1.4\times10^{-3}$ c s$^{-1}$ in
the 0.7 -- 10 keV band, based on the combined data of GIS2 and GIS3
within a radius of $3'$ from the source position.  Assuming a
Raymond-Smith model with temperature $kT = 3.0$ keV, metal abundance
$Z=0.3$, and $N_{\rm H} = 1 \times 10^{21-24}\ {\rm cm}^{-2}$, the
95 \% upper limit of the $N_{\rm H}$ corrected luminosity in a 0.5--10 keV band
is estimated to
be $1\times10^{30}-2\times 10^{32}\ {\rm erg\ s^{-1}}$ assuming the
distance to be 440 pc as shown in table 1.

The ROSAT All Sky Survey (RASS) observation covered this region in Jan
8--18, 1991, with an exposure of 877 s and showed no significant X-ray
emission.  The upper-limit intensity in the energy range 0.1--2.0 keV
is 0.02 c s$^{-1}$ at the $3\sigma$ confidence level.


The Class I protostar in the L1152 dark cloud, IRAS 20353+6742, was
located in the southwest direction in the GIS field of view (indicated
by an open circle in figure 1), but the source again showed no
significant X-ray emission.
Although the source position falls on the built-in calibration source
($^{55}$Fe) in GIS3, and also close to the edge of the GIS field of
view, we tried to estimate the upper limit of the luminosity of IRAS
20353+6742 in a similar way as for L1157-mm (table 1).  The upper
limit luminosity is about three times higher than that of L1157-mm because of the large
offset angle.

\subsection{Other Sources}


Besides the non detection of L1157-mm and IRAS 20353+6742, several X-ray sources are
clearly seen in figure 1.  We performed a sliding cell method to pick
up the sources. The cell size was $3'\times 3'$ (close to half power
diameter of the XRT $\sim3'$), and the soft(0.7--2.0 keV) and
hard(2--10 keV) band images were separately examined. We set a $5
\sigma$ threshold for the intensity in the a circle of $3'$ diameter,
and 9 sources were significantly detected and listed in table 2. This
table shows the significance for a diameter of $3'$ which contains about
50\% of the photons from a point source.

Among the 9 sources, 3 objects (\#6, \#8, and \#9) are detected in both
energy bands, and 6 objects are detected in either soft or hard band
only. Sources \#2, \#3, and \#4 are soft-band detection, and \#1, \#5, and
\#7 are hard-band detection, respectively. Four sources, \#1, \#4, \#6, 
and \#9, have no cataloged counterpart in NED, SIMBAD database, and TYCHO catalog(Egret et al.\ 1992) within our position accuracy, $r
< 1'$, and they are considered as new detections. We will call them as AXJ
2035+6806 (\#1), AXJ 2036+6800 (\#4), AXJ 2038+6801 (\#6), and AXJ 2041+6800
 (\#9), respectively.  Three sources, \#2, \#3, and \#8, were already
detected in RASS observation, and \#2 and \#8 have been identified as 
BWE 2034+6753 and HD 197471, respectively, within the position error of
the ROSAT PSPC\@. The remaining two sources, \#5 and \#7, are difficult
to judge whether they are really X-ray sources because of the poor
statistics. We note that the source \#5 is close to NGC 7023 SKCM 11; a
star of spectral type F2V at a distance of 240 pc (Strai$\check{\rm
z}$ys et al.\ 1992). X-ray properties of the detected sources excluding
\#5 and \#7 are described below.

\subsubsection*{\#1, and \#2, \#3 --- AXJ 2035+6806 and RASS sources }

The position of source \#1 is between \#2 and \#3
which are detected in the RASS observation, and the offset angles 
from these sources are $2.1'$ and $2.5'$, respectively. 
The position of \#2 is in agreement with the coordinate of BWE
2034+6753. Since these 2 sources are not detected in the GIS hard
band, their spectra are thought to be considerably soft. However, the
poor statistics and the coupling with \#1 hamper detailed
spectral study of these sources. In contrast, source \#1 emits X-rays
only in the hard energy band.  The fluxes of \#1, \#2, and \#3 in table 2
could be overestimated because of the mutual flux contamination.

\subsubsection*{\#4 --- AXJ 2036+6800}

This source is detected at $9'$ west from the field center. The
coordinate is (R.A., Decl.)$_{\rm J2000}$ = (${\rm 20^h36^m26^s},
68^\circ0'11''$) as determined from the center of gravity in the GIS
image.  The general accuracy of the position determination with ASCA
has been studied for SIS data, which give better position resolution
than GIS ones. The accuracy with the SIS is $40''$ at the 90\%
confidence level (Gotthelf 1996), which is mostly determined by the
systematic error in the satellite attitude calculation. Considering
the GIS pixel size of $15''$, the positional accuracy of the new
source is estimated to be $\sim 1'$. The RASS observation did not
indicate significant X-ray emission at this coordinate.  

This source is rather faint and very soft in contrast to the next AXJ
2038+6801. The GIS hard-band image (figure 1(b)) shows no X-ray emission
at the peak position of the soft-band image (figure 1(a)).  Figure 2(a)
shows the background-subtracted spectrum for the combined GIS2 and GIS3
data fitted with an absorbed thermal bremsstrahlung model.  The GIS
blank sky data are used as the background. The best-fit spectral
parameters are $kT = 0.27^{+0.47}_{-0.14}$ keV, $N_{\rm H} =
1.1^{+11}_{-1.1}\times 10^{21}$ cm$^{-2}$ (the error shows 90 \%
confidence level) shown in table 3, and the absorbed flux in 0.5--2.0
keV band is $\sim 1 \times 10^{-13}\ {\rm erg\ cm}^{-2}\ {\rm s}^{-1}$.
The closest object to this source is a star BD+67 1258 which is
identified as NGC 7023 SCKM 8; a G0IV star with $m_{\rm V}\sim 10.30$,
at a distance of 180 pc (Strai\v{z}ys et al.\ 1992). The angular
distance of this candidate star, $1.2'$, is slightly larger than the
positional error.


\subsubsection*{\#6 --- AXJ 2038+6801}

AXJ 2038+6801 is clearly detected in the center of the GIS image as
shown with a filled triangle in figure 1.  The spatial extent of the
source is consistent with the XRT point spread function. The
coordinate is (R.A., Decl.)$_{\rm J2000}$ = (${\rm 20^h37^m59^s},
68^\circ1'2''$), and the position accuracy is estimated to be the
same as the previous case, i.e.\ $\sim1'$.  Again, there is no X-ray
emission in the RASS observation.  The counting
rate in the 0.5--10 keV band is 0.015 c s$^{-1}$ for the sum of GIS2
and GIS3 within a radius of $6'$.

The spectrum is shown in figure 2(b) fitted with an absorbed power-law model. 
Spectral models for thermal bremsstrahlung and power-law emission are 
both acceptable with appropriate absorption. The resultant parameters are 
shown in table 3. The bremsstrahlung fit
gives a temperature $kT=7.6^{+16}_{-3.5}$ keV, and the power-law
model indicates a photon index to be $2.0 \pm 0.5$.  For both models,
the observed 0.5--10 keV flux is $5.8 \times 10^{-13}$ erg cm$^{-2}$
s$^{-1}$ and the absorption column density $N_{\rm H}$ is as large as
$ 2\times10^{22}\ {\rm cm}^{-2}$ which is significantly larger than
the Galactic interstellar absorption of $ 1.4\times10^{21}$ cm$^{-2}$
based on the 21cm line measurement (Dickey, Lockman 1990).  The
absorbed GIS spectrum predicts a PSPC count rate of 0.0025 c s$^{-1}$,
which is lower than the actual $3\sigma$ upper limit (0.02 c s$^{-1}$)
for this position. It is, therefore, natural that the source is
undetected with PSPC\@. We added a Gaussian line to the thermal
bremsstrahlung model at a center energy of 6.4 and 6.7 keV assuming a
neutral and He-like iron-K line, respectively.  The upper limit of the
equivalent width is 708 eV (6.4 keV line) and 1020 eV (6.7 keV line)
at the 95\% confidence level, respectively.

We also performed time variability analysis to search for flare like
events or pulsations.  The light curve shows no significant
variability during the total observation interval $\sim 100$ ks.  The
highest counting rate for a bin size of an hour is about twice the
average, which is typically 207 counts for the sum of GIS2
and GIS3 within a circle of $r=6'$ in 0.5--10 keV and dominated by
the source flux.  Also, a power spectrum obtained from an FFT analysis
(Leahy et al.\ 1983) does not show any significant peak above the 99\%
confidence level over a frequency range $3.9 \times10^{-5}$ Hz 
-- 16 Hz. The largest power appears at a
frequency 1.4 Hz, and the upper limit for a relative amplitude is 49\%
of the intensity at the 99\% confidence level assuming a sinusoidal
pulse shape.

\subsubsection*{\#8 --- HD 197471}

The peak position of this source is consistent with that of HD 197471
within about $0.5'$, which is also identified as NGC 7023 SCKM 23; a
star with a spectral type F7V at a distance of 94 pc (Strai\v{z}ys et
al.\ 1992).  The RASS data shows an X-ray source at the same position.
Figure 2(c) shows the observed GIS spectrum fitted with an absorbed
thermal bremsstrahlung model. The best-fit spectral parameters are $kT
= 0.66^{+1.4}_{-0.44}$ keV, $N_{\rm H} = 1.2^{+2.6}_{-1.0}\times 10^{22}$
cm$^{-2}$ (the error shows the 90 \% confidence level), and the
absorbed flux in 0.5--2.0 keV band is $\sim 2 \times 10^{-13}$ erg
cm$^{-2}$ s$^{-1}$. The absorption corrected luminosity is estimated
to be $1.2 \times 10^{30}$ erg s$^{-1}$ for the distance of 94 pc. The
positional consistency suggests that this source is the star HD
197471, emitting a spectrum with $kT\sim 0.7$ keV\@.

\subsubsection*{\#9 --- AXJ 2041+6800}
This source is not identified,
and no corresponding source is seen in the RASS data. The observed
spectrum is shown in figure 2(d) fitted with an absorbed power-law
model including a gaussian line. The 0.5--10 keV flux is $\sim 4
\times 10^{-13}$ erg cm$^{-2}$ s$^{-1}$.  The reduced $\chi^2$ value
improves by the inclusion of a gaussian line with an equivalent width EW = 1.9 keV in the absorbed power-law
model (from 1.17 to 0.99) as shown in table 3. The reduction of
$\Delta \chi^2 = 5.5$ with the decreased degrees of freedom of 2
indicates that the line is significant at the 90\% confidence level
(Malina et al.\ 1976). The best-fit value of the line center energy is
$\sim 4.6$ keV, so that the implied redshift is $\sim 0.4$ assuming
that the line is either neutral or He-like iron-K line. Taking this
redshift and assuming $q_{0}=0.5$, the 0.5--10 keV luminosity at the source is estimated to be
$1.2 \times 10^{44} h^{-2}$ erg s$^{-1}$, where $h=H_0/100\ {\rm km\
s^{-1}\ Mpc^{-1}}$ is the dimensionless Hubble constant.
We also fit the spectrum with an absorbed Raymond--Smith model.
The best-fit parameters are temperature of $kT\sim 5.4$ keV, 
metal abundance of $Z\sim 2.2$ solar, and redshift of $z\sim 0.39$, respectively, as shown in table 3. Therefore, an unusually high metal abundance is implied 
if the emission line is real.

\section{Discussion}

The upper limit we have obtained for the X-ray luminosity of L1157-mm,
$10^{30-32}$ erg s$^{-1}$, is comparable to or less than those
of Class I objects, such as WL6 and EL29 in the
$\rho$ Ophiuchi dark cloud (Kamata et al. 1997). The Class I protostar
in the L1152 dark cloud, IRAS 20353+6742, showed no significant X-ray emission, either.

The X-ray luminosities of the Class I objects so far studied do not
show clear correlation with physical parameters such as $M_{\rm env},
L_{\rm bol}$, and $F_{\rm CO}$. A relative configuration between the
direction of the CO bipolar outflow and the line of sight is suggested
as a relevant parameter.  For IRAS 20353+6742, both $L_{\rm bol}$
($=3.3L_{\solar}$) and $M_{\rm env}$ ($=0.05 M_{\solar}$) take
intermediate values between those for WL6 ($L_{\rm
bol}=2.0L_{\solar}$ and $M_{\rm env}=0.02 M_{\solar}$) and EL29
($L_{\rm bol}=41.0L_{\solar}$ and $M_{\rm env}=0.09 M_{\solar}$;
Bontemps et al.\ 1996). In the $\rho$ Ophiuchi dark cloud, all the
inclination angles of X-ray emitting protostars are less than
$30^{\circ}$ (Sekimoto et al.\ 1997). In contrast, the CO outflow of
L1152 shows almost an edge-on configuration (Bontemps et al.\ 1996).
Since an absorbing gas with $N_{\rm H} \gtsim 10^{23}$ cm$^{-2}$ can
hide an object with $L_{\rm X} \sim 10^{31}$ erg s$^{-1}$ as indicated
in table 1, the present negative result for IRAS 20353+6742 can be understood in
terms of a geometrical effect.

So far, no significant X-ray emission from Class 0 protostars has been
detected including L1157-mm.  We may consider three possible causes of
the low X-ray luminosity of Class 0 objects: 1) intrinsic low activity
for the X-ray emission in this class, 2) absorption by thick
surrounding gas or dust, 3) some effect due to source configuration.
In the first case, Class 0 objects are in the evolutionary stage where
the X-ray emission has not started yet.  In the second case, the thick
disk component ($N_{\rm H}>10^{23-24} {\rm cm}^{-2} $) hampers X-ray
detection of an object with $L_{\rm X} \sim 10^{31}$ erg s$^{-1}$ in
the present energy range (see table 1).  In fact the envelope mass of
L1157-mm ($\sim 3 M_{\solar}$) is significantly larger than those of
WL6 and EL29, and the envelope is supposed to extend around the
core. In the last case, the lack of clear X-ray emission from L1157-mm
can be understood because the inclination angle of CO outflow is
$\sim80^\circ$, i.e.\ an almost edge-on configuration, assuming that the X-ray
emission of protostars is collimated parallel to the CO outflow.  If
this is the case, there remains a possibility for detecting X-ray
emission from Class 0 protostars for which outflows are looked at in
the exactly pole-on configuration (with an inclination angle of $0 \pm
10^{\circ}$), even if the envelope mass is larger than those of Class
I protostars.

The most luminous new X-ray source in the field, AXJ 2038+6801, shows a hard X-ray spectrum with
strong absorption and no obvious time variability. Within $5'$ from
the source, no identification is made with SIMBAD, NED database, and
TYCHO catalog (Egret et al.\ 1992), which leaves various possibilities
for the source nature.  Assuming the distance of AXJ 2038+6801 to be
the same as that of the L1157 cloud (440 pc), the 0.5--10 keV luminosity
corrected for the absorption is estimated to be $\sim2\times10^{31}\ {\rm
  erg\ s}^{-1}$. This is similar to the luminosity of X-ray luminous
young stellar objects.  The ratio of the X-ray measured $N_{\rm H}$ to
the optically estimated $A_{\rm V}$ for L1157-mm (20 mag: Davis,
Eisl\"offel 1995) suggests a relation $\sim 10^{21}\ {\rm cm}^{-2}\ 
{\rm mag}^{-1}$: which is consistent with those for X-ray luminous
Class I objects.  Hence, if L1157 dark cloud has a large spatial
extent ($\sim 0.8$ pc) and reaches the position of AXJ 2038+6801, the
observed strong absorption is naturally explained.

If AXJ 2038+6801 is not associated with L1157 and is located within
our galaxy, the luminosity has to be roughly in the range $10^{31} -
10^{34}\ {\rm erg\ s}^{-1}$. Galactic X-ray binary sources containing
neutron stars or black holes generally exhibit hard X-ray spectra, but
their luminosities are mostly higher by orders of magnitude. White
dwarf binaries and Crab-like pulars have X-ray luminosities in this
range and produce fairly hard spectra. The present negative detection
of X-ray pulses suggest that the possibility for Crab-like pulsars is
low unless the pulse frequency is faster than 16 Hz. Dwarf novas show
X-ray luminosities in the range $10^{31-33}$ erg s$^{-1}$ and hard
spectra with $kT \gtsim 10$ keV\@. Only the lack of time variation is
the different characteristics of AXJ 2038+6801 from dwarf
novas. Therefore, this possibility remains as the nature of AXJ
2038+6801.  The hard spectrum of this source also suggests a
possibility of intrinsically absorbed or obscured AGN\@. Based on the
$\log N-\log S$ relation (Ueda et al.\ 1999), it is expected that one
or two serendipitous sources with this flux level ($\sim 5 \times
10^{-13}$ erg cm$^{-2}$ s$^{-1}$) are detected in the GIS field of
view in a 60 ks observation.

Another new source AXJ 2036+6800 shows a significantly softer
spectrum, with no detectable X-ray emission above 2 keV\@. The
absorption is much weaker than the previous case of AXJ
2038+6801. Regarding its spectrum ($kT \sim 0.3$ keV), we will
consider a possibility of a late type star. Assuming that the source
has the same distance as the closest star BD+67 1258 (spectral type
G0IV), the 0.5--2.0 keV luminosity becomes $8.0\times10^{29}\ {\rm erg\
s}^{-1}$. This is in the higher end in the luminosity distribution of
main-sequense (F and G type) stars with $kT\sim 0.3$ keV (Schmitt et
al. 1990). It is, therefore, possible that the source can be a
late-type star. 

The spectrum of AXJ 2041+6800 shows an interesting feature which may
be a redshifted iron line. A simple $\chi^2$ test indicates that the
line is significant at the 90\% confidence as shown in the previous
section. However, the line energy is close to the L-edge energy (4.78
keV) of Xe, which is the detector gas of the GIS instrument. We should
be cautious about the detection of the strong iron line, and should
wait for further observations with better statistics. If the redshift
is really around 0.4, the implied luminosity indicates that the source
is either an AGN or a cluster of galaxies.

Among the other detected sources, three have been detected with the
RASS\@. One is identified to be an F type star, HD 197471, and its
spectrum ($kT < 1$ keV) and luminosity ($1.2 \times 10^{30}$ erg
s$^{-1}$ in 0.5--2.0 keV) are consistent with the emission from a
late-type star.  The other two sources are close to each other and
confused with AXJ 2035+6806. Observations with higher angular
resolution are needed to look into the nature of these sources.

\vspace{1pc}\par
The authors thank Y. Sekimoto, M. Umemoto for providing the recent
radio observation results, S. Sasaki for discussion, and Y.
Ishisaki, K. Kikuchi, and Y. Ueda for helping the data analysis.
We also thank the anonymous referee for useful comments and suggestions.
NASA/IPAC Extragalactic Database (NED) is operated by Jet Propulsion
Laboratory, Caltech, under contract with NASA and the SIMBAD data base
is operated by the Centre de Donnes astronomiques de Strasbourg.

\section*{References}
\re Andr\'e P., Ward-Thompson D., Barsony M.\ 1993, ApJ 406, 122
\re Bachiller R.\ 1996, ARA\&A 34, 111
\re Bontemps S., Andr\'{e} P., Terebey S., Cabrit S.\ 1996, A\&A 
 311, 858
\re Burke B.E., Mountain R.W., Harrison D.C., Bauts M.W., Doty J.P.,
 Ricker G.R., Daniels P.J.\ 1991, IEEE Trans., ED-38, 1069
\re Casanova S., Montmerle T., Feigelson E.D., Andr\'e P.\ 1995, 
 ApJ 439, 752
\re Davis C.J., Eisl\"offel J.\ 1995, A\&A 300, 851
\re Dickey J.M., Lockman F.J. 1990, Ann. Rev. Ast. Astr. 28, 215
\re Egret D., Didelon P., McLean B.J., Russell J.L., Turon C.\ 1992, A\&A 258, 217
\re Gotthelf E.\ 1996, {\it ASCA} News 4, 31
\re Gueth F., Guilloteau S., Dutrey A., Bachiller R.\ 1997, A\&A 323, 943
\re Kamata Y., Koyama K., Tsuboi Y., Yamauchi S.\ 1997, PASJ 49, 461
\re Koyama K., Hamaguchi K., Ueno S., Kobayashi N., Feigelson E.D.\ 1996, PASJ
 48, L87
\re Leahy D.A., Darbro W., Elsner R.F., Weisskopf M.C., Sutherland P.G., Kahn S., Grindlay J.E.\ 1983, ApJ 266, 160
\re Malina R., Lampton M., Bowyer S. 1976, ApJ 209, 678
\re Myers P.C., Fuller G.A., Mathieu R.D., Beichman C.A., Benson P.J., 
 Schild R.E., Emerson J.P.\  1987, ApJ 319, 340
\re Ohashi T., Ebisawa K. Fukazawa Y., Hiyoshi K., Horii M., Ikebe Y., 
 Ikeda H., Inoue H. et al.\ 1996, PASJ 48, 157
\re Ozawa H., Nagase F., Ueda Y., Dotani T., Ishida M.\ 1999, ApJ 523, L81
\re Sekimoto Y., Tatematsu K., Umemoto T., Koyama K., Tsuboi Y., Hirano N.,
 Yamamoto S.\ 1997, ApJ 489, L63
\re Schmitt J.H.M.M., Collura A., Sciortino S., Vaiana G.S., Harnden
F.R.\ Jr,  Rosner R.\ 1990, ApJ 365, 704
\re Strai$\check{\rm z}$ys V., $\check{\rm C}$ernis K., Kazlauskas A.,
Mei$\check{\rm s}$tas E.\ 1992, Baltic Astronomy 1, 149
\re Tafalla M., Bachiller R.\ 1995, ApJ 443, L37
\re Tanaka Y., Inoue H., Holt S.S.\ 1994, PASJ 46, L37
\re Ueda Y., Takahashi T., Inoue H., Tsuru T., Sakano M., Ishisaki Y.,
 Ogasaka Y., Makishima K., Yamada T., Akiyama M., Ohta K.\ 1999,
 ApJ 518, 656
\re Umemoto T., Iwata T., Fukui Y., Mikami H., Yamamoto S., Kameya O., 
Hirano N.\ 1992, ApJ 392, L83

\begin{table*}[htb]
\begin{center}
Table~1.\hspace{4pt}The 95\%-confidence upper limit for the X-ray
luminosity of L1157-mm and IRAS 20353+6742 after correction for the
absorption in the energy band 0.5--10 keV\@.
\end{center}
\begin{tabular*}{\textwidth}{@{\hspace{\tabcolsep}
    \extracolsep{\fill}}ccc}\hline\hline\\[-6pt]
$N_{\rm H}$ &  $L_{\rm x}$ (0.5--10 keV) & $L_{\rm x}$ (0.5--10 keV) \\
(cm$^{-2}$) & ($10^{31}$ erg s$^{-1}$ ) & ($10^{31}$ erg s$^{-1}$ ) \\
& L1157-mm & IRAS 20353+6742 \\\hline
 $1\times10^{21}$ & $< 0.13$ & $<0.39$\\
 $1\times10^{22}$ & $< 0.22$ & $<0.67$\\
 $1\times10^{23}$ & $< 1.1$  & $<3.5$\\
 $1\times10^{24}$ & $< 20$  & $<77$\\\hline
\end{tabular*}\\
\end{table*}

\clearpage
\begin{table*}[htb]
\begin{center}
  Table~2.\hspace{4pt} The source list detected in the GIS field with
  more than $5 \sigma$ significance. Sources \#6, \#8, and \#9 are
  detected in both soft (0.7--2.0 keV) and hard (2.0--10 keV) energy
  bands. Souces \#2, \#3, and \#4 are detected in the soft band only,
  and \#1, \#5, and \#7 are in the hard band only.
\end{center}
{\footnotesize
\begin{tabular*}{\textwidth}{@{\hspace{\tabcolsep}
    \extracolsep{\fill}}cllcccl}\hline\hline\\[-6pt]
No. & \multicolumn{1}{c}{R.A.}& \multicolumn{1}{c}{Decl.} &
\multicolumn{2}{c}{Significance ($\sigma$)$^\ast$}  & $F_{\rm X\ (0.5-10~keV)}$ & ID$^{\dagger}$ \\
    & \multicolumn{1}{c}{J2000} & \multicolumn{1}{c}{J2000} & soft &
hard & ($10^{-13}$ erg cm$^{-2}$ s$^{-1}$) & \\\hline
\#1 &$20^h35^m19^s$ & $68^\circ06'19''$ & 4.70 & 7.35 & $3.7$ & not available (AXJ 2035+6806)\\
\#2 &$20^h35^m25^s$ & $68^\circ04'51''$ & 5.54 & 4.19 & $3.8$ & BWE 2034+6753, RASS source \\
\#3 &$20^h35^m33^s$ & $68^\circ07'34''$ & 7.05 & 4.08 & $3.8$ & RASS source \\
\#4 &$20^h36^m26^s$ & $68^\circ00'11''$ & 7.49 & 2.71 & $1.0$ & BD+67 1258 ? (AXJ 2036+6800) \\
\#5 &$20^h37^m34^s$ & $67^\circ55'06''$ & 2.15 & 5.89 & $0.5$ & NGC 7023 SCKM 11 \\
\#6 &$20^h37^m59^s$ & $68^\circ01'02''$ & 10.0 & 21.9 & $5.8$ & not available (AXJ 2038+6801) \\
\#7 &$20^h38^m55^s$ & $68^\circ07'22''$ & 1.93 & 5.35 & $1.4$ & not available\\
\#8 &$20^h40^m02^s$ & $67^\circ47'56''$ & 10.5 & 5.53 & $2.0$ & HD 197471, RASS source \\
\#9 &$20^h41^m04^s$ & $68^\circ00'09''$ & 8.02 & 6.43 & $3.5$ & not available (AXJ 2041+6800) \\
\hline
\end{tabular*}}\\
\vspace{6pt}\par\noindent
$\ast$ Estimated within a diameter of $3'$.
\par\noindent
$\dagger$ Names in parentheses are used in the text.
\end{table*}

\begin{table*}[htb]
\begin{center}
  Table~3.\hspace{4pt} Spectral parameters for source \#4, \#6, \#8, and
  \#9. The errors are statistical and 90\%-confidence levels for single
  parameter of interest.
\end{center}
{\footnotesize
\begin{tabular*}{\textwidth}{@{\hspace{\tabcolsep}
    \extracolsep{\fill}}lllllll}\hline\hline\\[-6pt]
   & Target name & Model   & $N_H$  & $\Gamma / kT$ & EW$^\dagger$/$Z^\ddagger$  &$\chi^2/d.o.f$ \\
    &&& ($10^{22}\ {\rm cm}^{-2}$) &  (  /keV) & (keV/solar) \\\hline 
    \#4& AXJ  2036+6800    & Bremsstrahlung$^\ast$  & $0.11^{+1.1}_{-0.11}$ 
    & $0.27^{+0.47}_{-0.14}$& & 5.14/10\\
    \#6& AXJ 2038+6801     & Power law$^\ast$       & $2.3^{+1.1}_{-1.0}$ 
    & $2.0^{+0.5}_{-0.5}$   & & 23.9/22\\
        &                   & Bremsstrahlung  & $1.9^{+0.8}_{-0.7}$ 
    & $7.6^{+16}_{-3.5}$    & & 23.4/22\\
    \#8& HD 197471     & Bremsstrahlung$^\ast$  & $1.2^{+2.6}_{-1.0}$ 
    & $0.66^{+1.4}_{-0.44}$ & & 5.03/8\\
    \#9& AXJ  2041+6800     & Power law       & $0.61^{+2.7}_{-0.61}$ 
    & $1.7^{+1.6}_{-0.78}$  & & 23.4/20\\
      &                   & Power law + Gaussian$^\ast$ &$0.85^{+3.0}_{-0.85}$ 
    & $2.2^{+1.8}_{-1.0}$   & $1.9^{+1.5}_{-1.3}$ $^\dagger$ & 17.9/18\\
        &                   & Raymond-Smith   & $0.68^{+2.1}_{-0.68}$ 
    & $5.4^{+6.7}_{-2.6}$   & $2.2^{+2.8}_{-1.9}$ $^\ddagger$ & 21.1/21\\
    \hline
  \end{tabular*}}\\
\vspace{6pt}\par\noindent
$\ast$ Models fitted to the data in figure 2.
\par\noindent
$\dagger$ Equivalent width.
\par\noindent
$\ddagger$ Metal abundance.\\
\end{table*}

\clearpage
\begin{fv}{1}{0.1cm}{The GIS image in two energy bands: 0.7--2.0 keV (a), 
    and 2.0--10 keV (b). Dashed circles show L1157 and L1152 regions
    with a radius of $2'$, which roughly corresponds to the size of CO
    outflows.  Filled triangles indicate positions of nine new X-ray
    sources detected in the GIS field (see table 2). Filled
    squares show positions of identified objects, BWE 2034+6753 (radio source), 
    BD+67 1258 (star), NGC 7023 SCKM 11 (star),
    and  HD 197471(star) for sources \#2, \#4, \#5, and
    \#8, respectively.  The surrounding dotted circle shows the
    effective field of view of GIS : $r=20'$.\\\vspace*{5mm}}
\\
    \psfig{figure=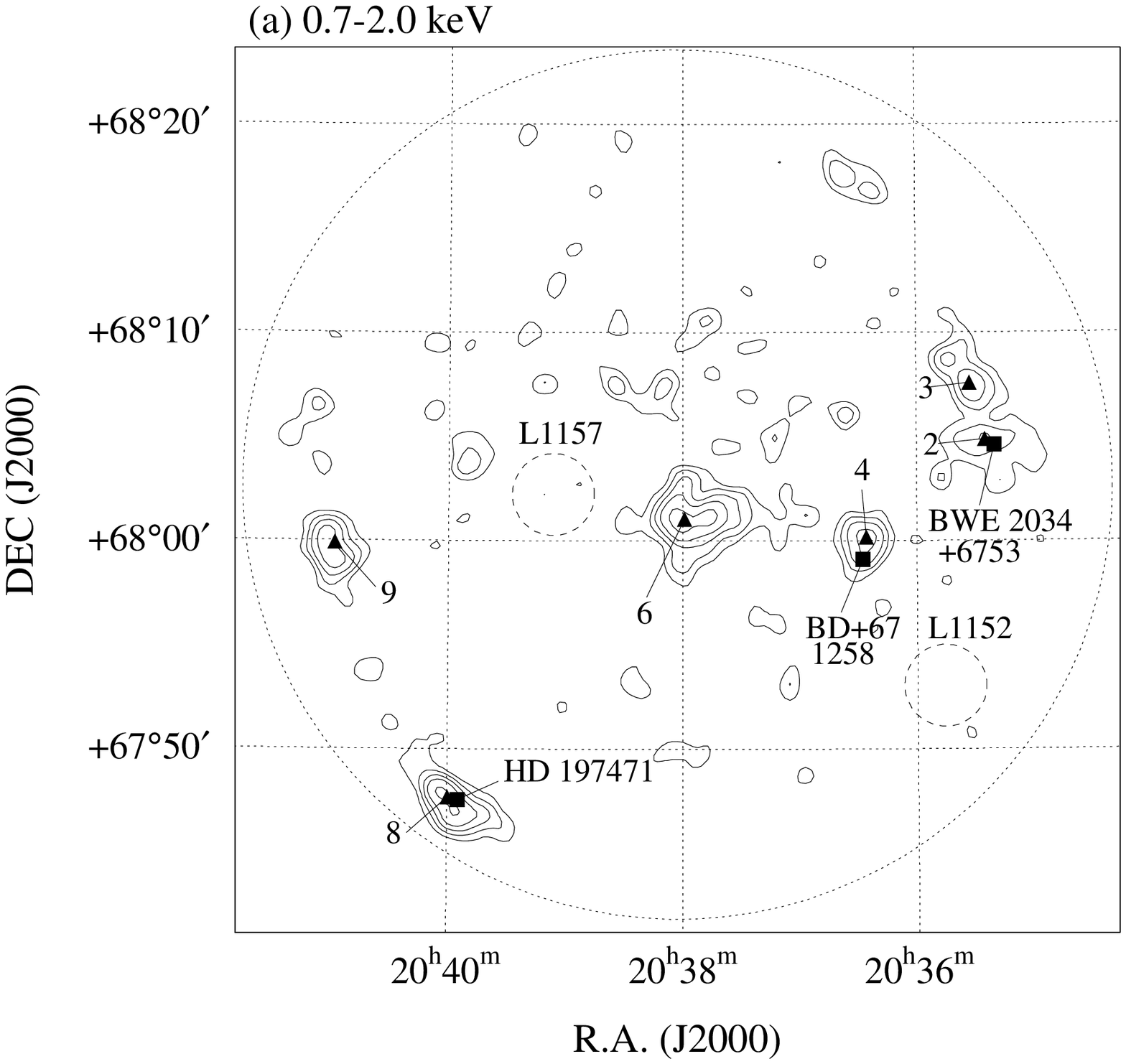,width=7cm}
    \psfig{figure=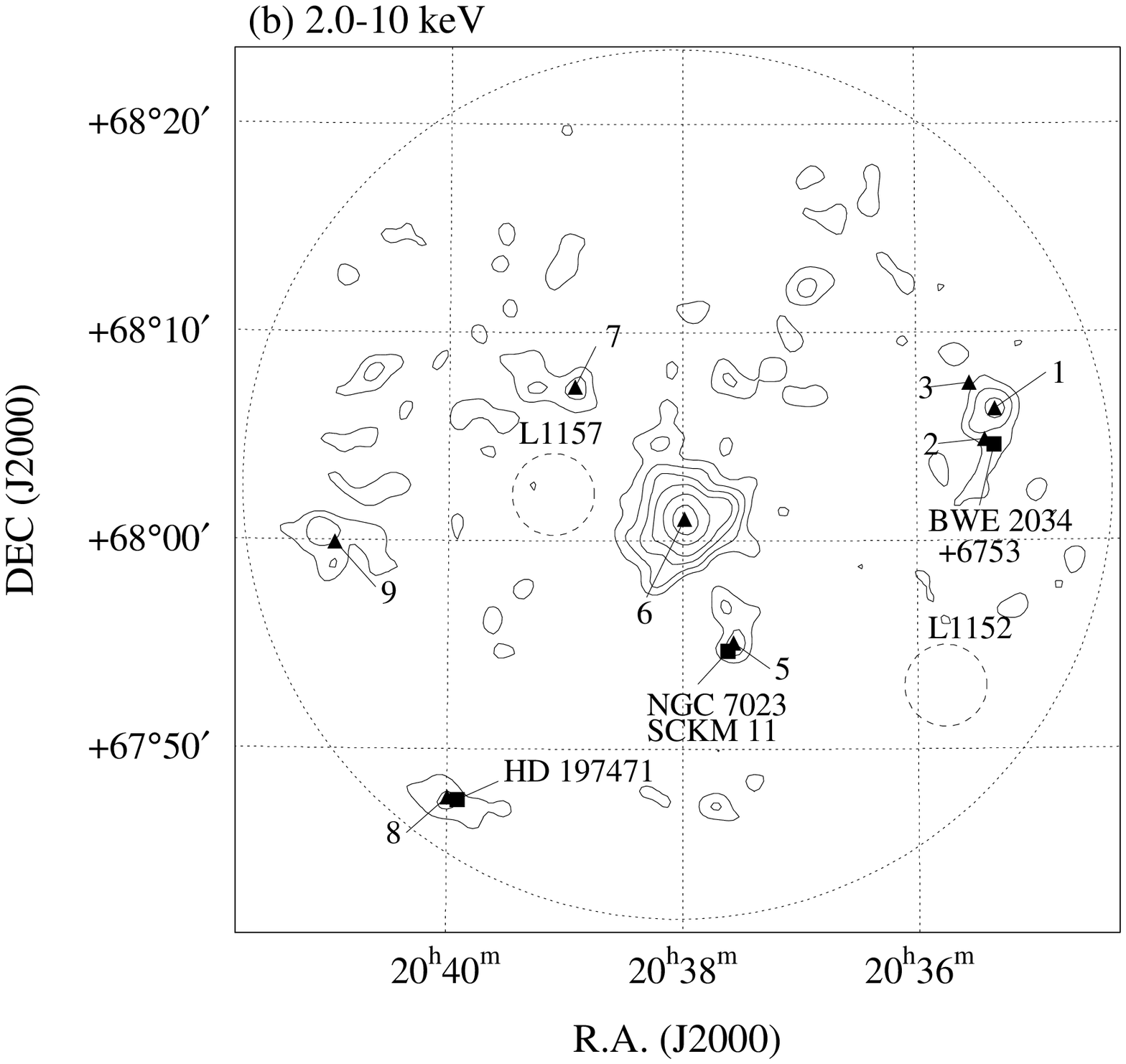,width=7cm}\\
\end{fv}
\clearpage

\begin{fv}{2}{0.1cm}{Spectra with GIS2 and 3 of new X-ray sources 
    AXJ 2036+6800(a), AXJ 2038+6801(b), HD 197471(c), and AXJ
    2041+6800(d).  The models fitted to the data are indicated with
    asterisks in table 3.  The bottom panels show residuals of the fits
    in units of standard deviation.\\\vspace*{5mm}}
\\
    \psfig{figure=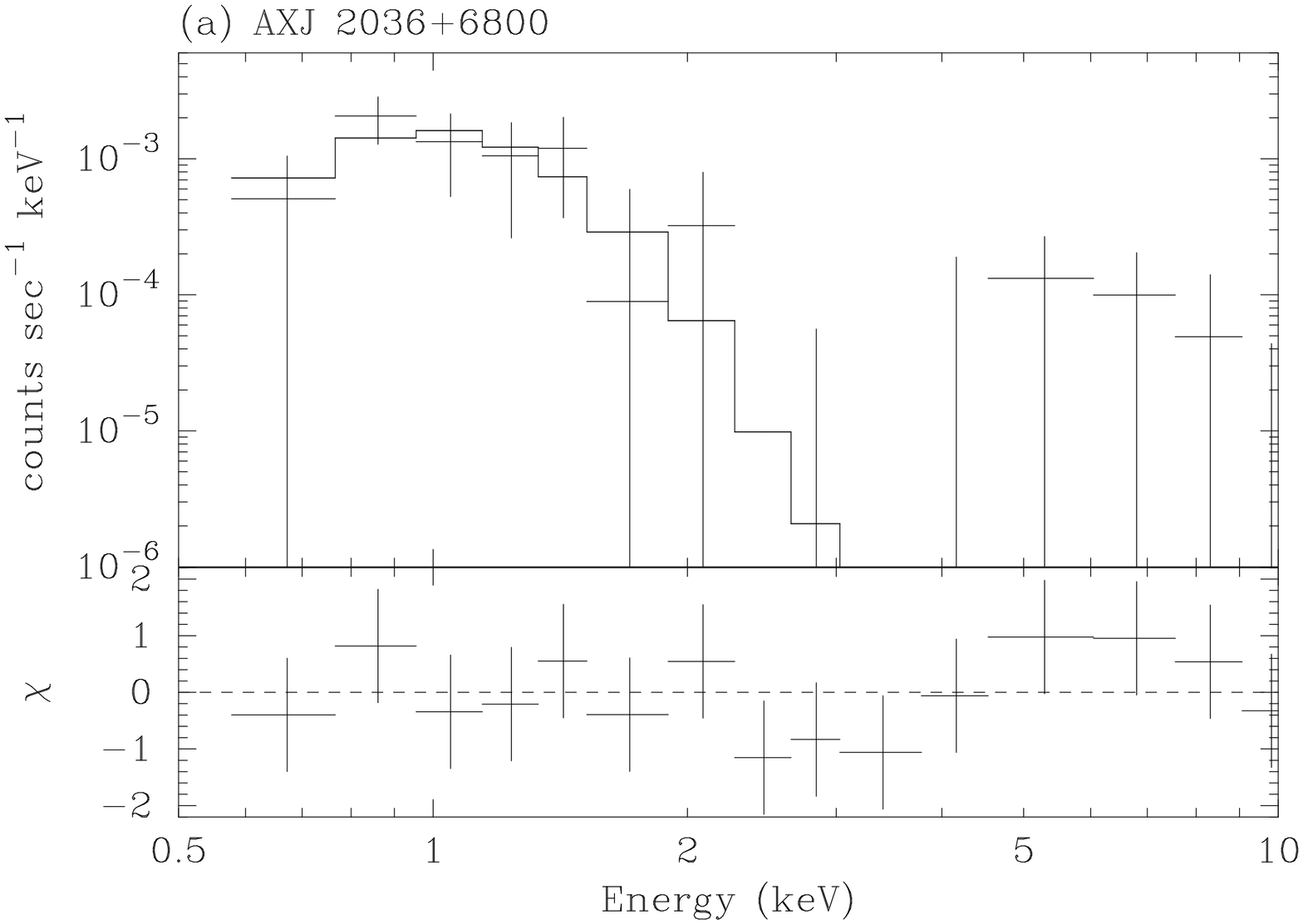,width=7cm,angle=-90}
    \psfig{figure=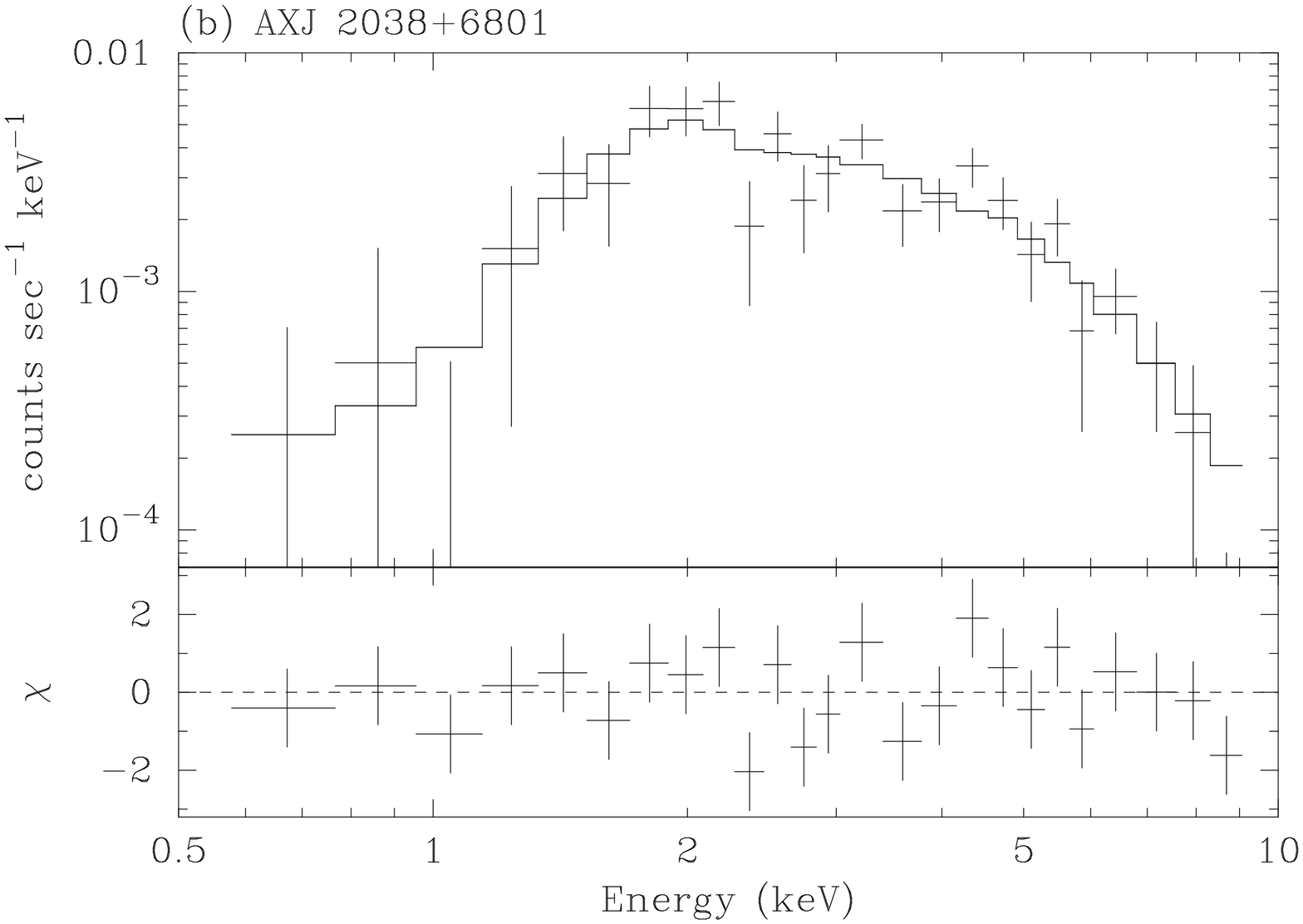,width=7cm,angle=-90}\\
    \psfig{figure=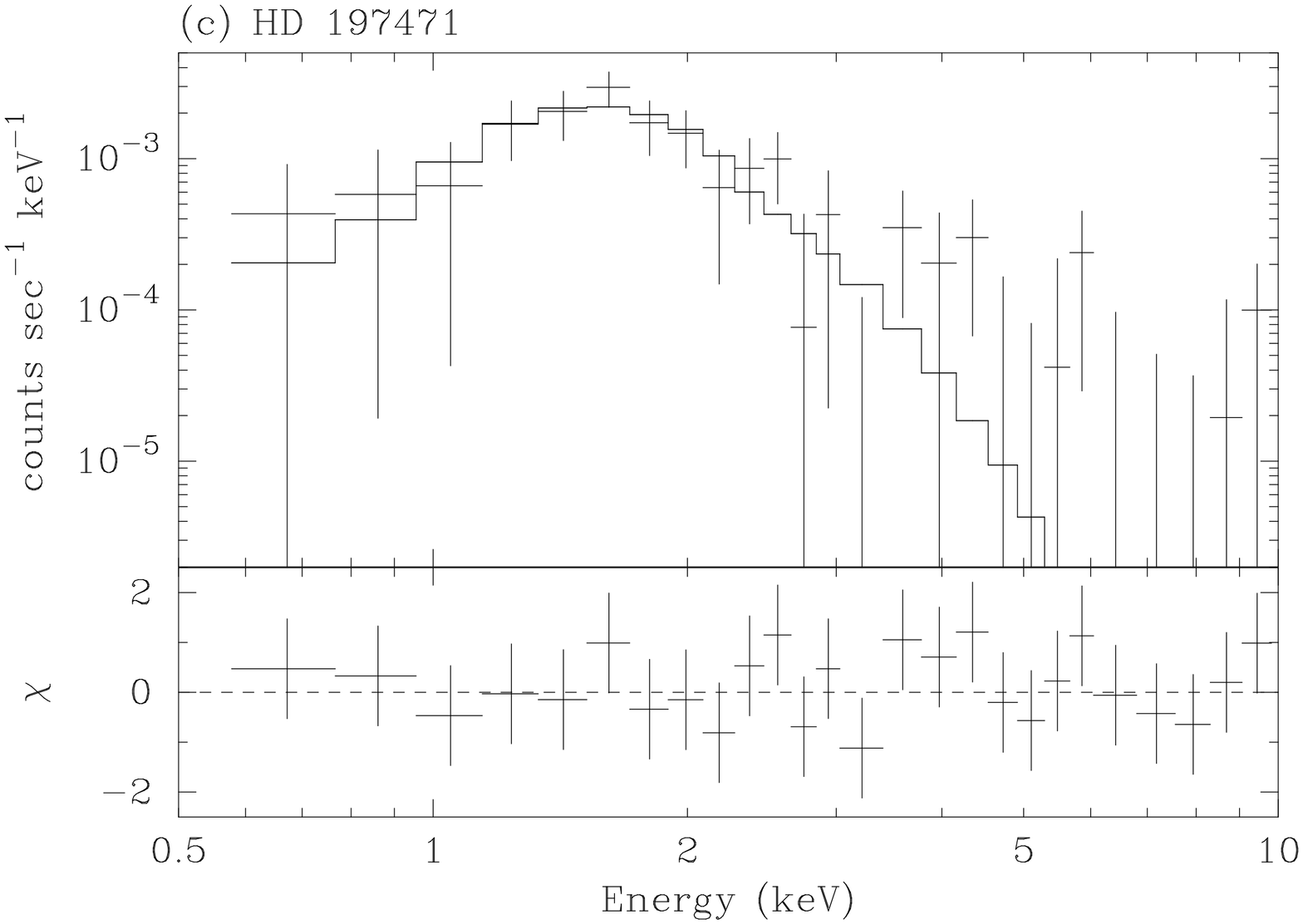,width=7cm,angle=-90}
    \psfig{figure=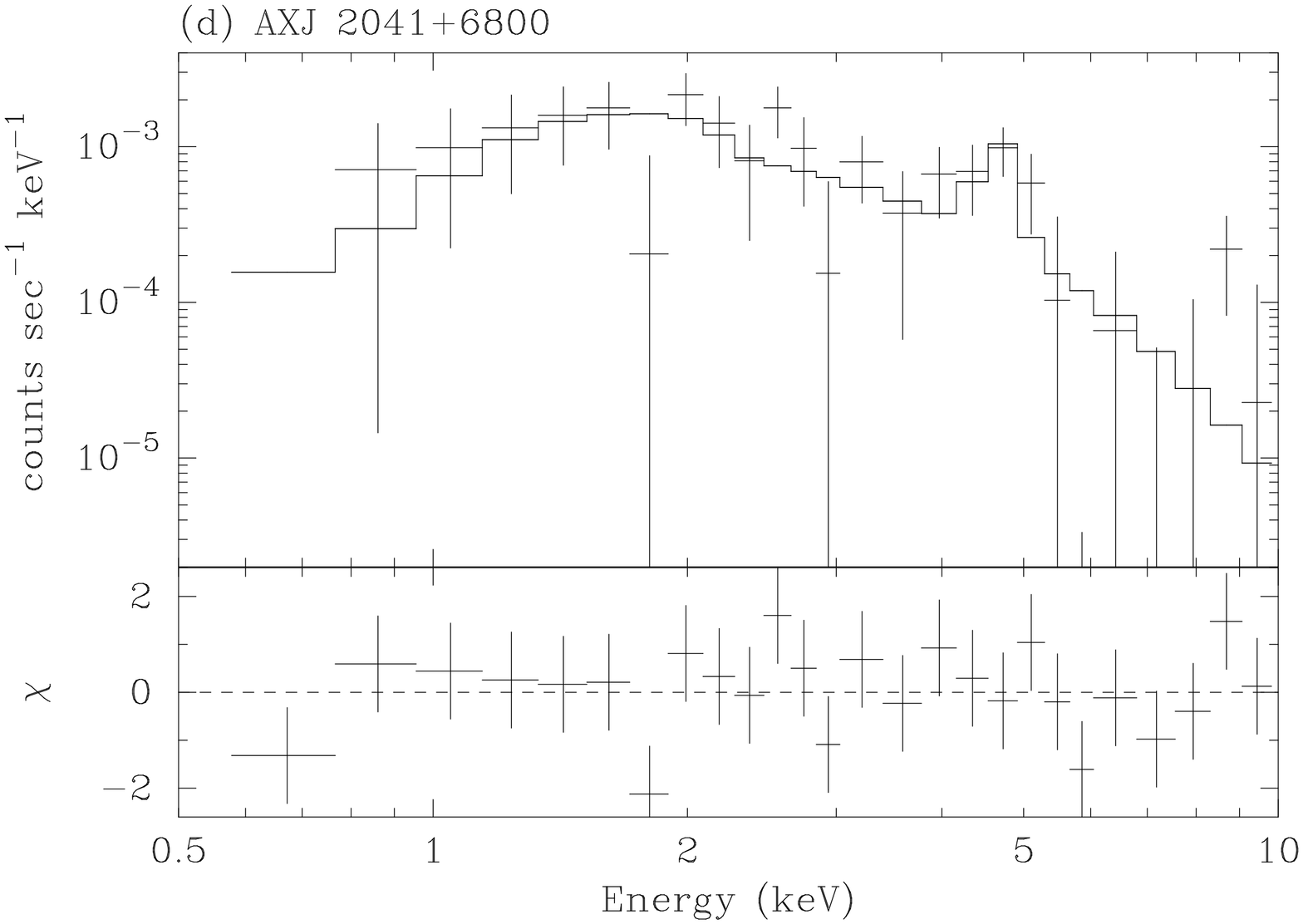,width=7cm,angle=-90}
\end{fv}
\end{document}